\def\pb{{\bf p}}\def\Ab{{\bf A}}\def\o{\omega}
\documentstyle[12pt]{article}
\begin{document}
\begin{titlepage}
\title{Comments on "Objectification of classical properties
induced by quantum vacuum fluctuations"}
\author{Lajos Di\'osi\thanks{E-mail: diosi@rmki.kfki.hu}\\
KFKI Research Institute for Particle and Nuclear Physics\\
H-1525 Budapest 114, POB 49, Hungary\\\\
{\it bulletin board ref.: gr-qc/9406041\hfill}}
\maketitle
\begin{abstract}
We argue that in standard quantum electrodynamics radiative corrections
do not lead to decoherence of unexcited atomic systems. The proposal
of Santos relies upon deliberate switching on and off the vacuum
interactions.
\end{abstract}
\end{titlepage}

In his recent Letter \cite{Santos},
Santos claims to show that "there exists a
{\it fundamental} mechanism producing decoherence of macroscopic bodies,
without any departure from quantum theory". This fundamental mechanism
is provided, as the Letter states, by the coupling of charged constituents
with the electromagnetic quantum vacuum.

We agree with the general part of the Letter's argument. Quantized
electromagnetic vacuum is, in many respects, similar to ordinary
reservoirs influencing the given quantum system and
destroying its coherence permanently \cite{JoosZeh,Zurek}. There is
a particular difference, however. The influence of vacuum must not lead
to any decoherence if the system is in unexcited state. We note that,
for the evolution of bare excited states, decoherence is a well-known
effect. The electromagnetic vacuum-fluctuations make the excited
state of an H-atom decay; the reduced density matrix of the atom will
be a mixture of the excited state and of the various ground states
corresponding to various recoils exerted by the emitted foton \cite{Diosi}.
The naiv idea that decoherence would occur for bare {\it ground}
states as well fails definitely. We think that the effect calculated
by Santos is due to the sudden switching on the vacuum influence at
$t=0$. Had we chosen the usual adiabatic
switching on we would get no decoherence at all.

We consider the sudden switching on at time $t=0$ unphysical.
In particular, it would excite ground state atoms and would make them radiate
after all.

Let us discuss the simple example presented by Santos.
Given an H-atom in bare ground state, switching on vacuum influence at $t=0$,
the dressed H-atom develops decoherence for $t$ big enough when
we trace over the fotonic degrees of freedom. The degree $\mu$ of
purity is calculeted and found to be different from unity (13).
Consider the Eqs.~(11) and (12) of the Letter.
Let us write down the explicit time dependence of the
needed matrix elements for the operators
$\pb$ and $\Ab$, respectively:
\begin{equation}
\langle\phi_n|\pb(t)|\phi_0\rangle=
\exp(i\o_n t)\langle\phi_n|\pb(0)|\phi_0\rangle~~,
\end{equation}
\begin{equation}
\langle k|\Ab(t)|0\rangle=\sum_k\exp(i\o t)\langle k|\Ab(0)|0\rangle
\end{equation}
where $\o$ is the energy of the one-foton state $|k\rangle$.
Let us calculate the matrix element of the Letter's
$U_1(t)$ in the limit $t\rightarrow\infty$.
Invoking our Eqs.~(1) and (2) we obtain:
\begin{equation}
\langle\phi_n|\langle k|U_1(\infty)|0\rangle|\phi_0\rangle=
        -i{e\over m}\sum_k\int\limits_0^\infty\exp\left[i(\o_n+\o)t\right]dt~
        \langle\phi_n|\pb|\phi_0\rangle\langle k|\Ab|0\rangle.
\end{equation}
This expression would vanish if the vacuum interaction  were switched
on adiabatically. In that case the time integral would be proportional
to \hbox{$\delta(\o_n+\o)$}
which is always zero since both $\o_n$ and $\o$ are
positive numbers. Hence, for adiabatic switching on, the degree
$\mu$ of purity remains equal to 1. The nontrivial result of Eq.~(13)
is a consequence of the sudden switching on at finite time $t$.

The ultimate argument against the sudden switching on is the following.
It is not at all accidental that, in the theory of quantum electrodynamics,
the interaction is switched on at $t=-\infty$.
Firstly, this choice is most
conform with reality where interaction has already been acting
since asymptotic past time. Secondly, a sudden switch
at any finite time would create
{\it real} particles even from the vacuum.
In the simple case shown in the Letter, the bare ground state of the
H-atom becomes excited as $U(t)$ is getting to act on it for $t>0$;
the excited state will then decay
and the emitted real fotons will be detectable.
The Born-approximation yields
\begin{equation}
{8\alpha\over3\pi m^2}\sum_n|\langle\phi_0|\pb|\phi_n\rangle|^2
        {\o\over(\o+\o_n)^2}
\end{equation}
for the spectral distribution of the emitted foton. The norm of this
distribution, i.e. the overall probability of foton emission by the
H-atom, is equal to $1-\mu$ which is the
degree of impurity of the final state.

Let us summarize our statement. In standard quantum electrodynamics
bare charges are getting dressed by {\it virtual} fotons.
As a matter of fact, the bare ground states such as, e.g.,
the bare ground states of atoms may change (Lamb-shift) but,
obviously, they never emit {\it real} fotons. To establish decoherence,
one {\it needs} real emitted fotons to trace out.
On one hand, the proposal \cite{Santos} has been based on the
principles of quantum electrodynamics. On the other hand, Santos has
replaced the usual adiabatic switching on the interaction, by a sudden one
at finite time, and this seems
to raise serious problems like, e.g., spontaneous radiation of ground states.
Furthermore,  the experimentally observed states are dressed states.
Actually, detectors are also based on electrodynamic interactions
and see the foton cloud of the charge in
question rather than the naked charge itself.
To observe decoherence of Santos, one has to detect the naked
charges inside their foton clouds.
This is possible only if we switch off again the
interaction with the radiation field which is obviously not the case
in Nature. Detectors, using alternative (i.e. not electrodynamic)
interactions could see the naked charges even behind their foton
clouds but such particular experiences of decoherence would not yield the
general objectification aimed by the Letter.

Our doubts follow from standard field-theoretical considerations.
We, of course, could not exclude that more refined and
less standard considerations would offer a plausible implementation of the
Letter's basic idea.

\bigskip
This work was supported by the grant OTKA No. 1822/1991.


\begin{thebibliography}{99}
\bibitem{Santos} E. Santos, Phys. Lett. A188, 198 (1994).
\bibitem{JoosZeh} E. Joos and H.D. Zeh, Z. Phys. B59, 223 (1985).
\bibitem{Zurek} W. H. Zurek, Phys. Today 44, 36 (1991).
\bibitem{Diosi} For an exact relativistic evolution equation of the reduced
atomic density operator see:
L. Di\'osi, Found. Phys. 20, 63 (1990).
\end{thebibliography}
\end{document}